\documentclass[letterpaper,12pt]{article}
\usepackage{tabularx} 
\usepackage{amsmath}  
\usepackage{braket}
\usepackage{physics}
\usepackage{mathrsfs}
\usepackage{bbm}
\usepackage{mathpazo, fourier}
\usepackage{qcircuit}
\usepackage{xypic}
\usepackage{calligra}
\usepackage{yfonts} 
\usepackage{mathrsfs}
\usepackage{siunitx}
\usepackage{cancel, xcolor, ulem}
\usepackage{graphicx} 
\usepackage[margin=1in,letterpaper]{geometry} 
\usepackage{cite} 
\usepackage[final]{hyperref} 
\hypersetup{
	colorlinks=true,       
	linkcolor=blue,        
	citecolor=blue,        
	filecolor=magenta,     
	urlcolor=blue         
}
\usepackage{blindtext}

\begin{document}

\title{Feynman's Entangled Paths to Optimized Circuit Design}
\author{Kartik Anand}
\date{\today}
\maketitle

\begin{abstract}
We motivate an intuitive way to think about quantum circuit optimization problem inspired by \textit{Feynman’s path formalism}. While the use of path integrals in quantum circuits remains largely underdeveloped due to lack of definition of the action functional for such systems. However this \textit{feynman's path} perspective leads us to consider about how entanglement evolution throughout the circuit can serve as a guiding principle for optimizing circuit design. We conjecture that an optimal \textit{state-path} is highly likely to belong to a family of \textit{paths} with the minimum possible \textit{path-entanglement sum}. This could enhance the efficiency of circuit optimization problems by narrowing the \textit{state-path} search space, leading to faster convergence and reliable outputs. Further, we discuss that for some special target states this conjecture may not provide significant insights to the circuit optimization problem and argue that such cases constitute only a small subset of the target sets encountered by a circuit optimization algorithm.
\end{abstract}

\section{Introduction}
Feynman's path integral \cite{feynmanpathintegral} provides an alternative to standard means of expressing quantum dynamics, much in the same way as lagrange formalism is an alternative way of expressing Hamiltonian dynamics. This \textit{"sum-over-paths" / "feynman's paths"} view involves calculating the transition probability amplitude between two states, whose transition rules are governed by a given Hamiltonian. In physics, \textit{feynman's formalism} provides an intuitive understanding of wavefunction at a space-time point $(x,t)$:
\begin{equation}
\psi(x, t) = \int K(x, t; x_0, t_0) \, \psi(x_0, t_0) \, dx_0
\end{equation}
where $K(x, t; x_0, t_0)$ is the \textit{Feynman propagator} with the initial \textit{space-time} coordinates $(x_0, t_0)$. From this expression, one can deduce that wavefunction at location $x$, at a later time $t$ is just the \textit{interference} of the \textit{waves} from all the sources ($x_0$) across the space. The amplitude of each wave at the \textit{space-time} point $(x,t)$ in space-time is governed by the \textit{Feynman propagator}.
    As noted by \cite{Baez1994}, \cite{Penney_2017}, \cite{rudiakgould2006sumoverhistoriesformulationquantumcomputing}, \cite{xu2021lagrangianformalismquantumcomputation},   Feynman's formalism can also be used to make work for discrete time systems, such as quantum circuits. \cite{Penney_2017} tried to formalize sum-over-paths formalism for a special kind of balanced circuits - Continuous Variable (CV) Clifford Circuits and Discrete Clifford Circuits with odd-prime qudits \textit{(quopits)} by determining symplectomorphism generating functions and using them as \textit{action functionals} in the transition amplitude expression. However, a transition amplitude expression for a generalized "two-qubit" Haar-random gates circuit is lacking in the literature. This makes it notoriously hard to use this formalism for any theoretical analysis in the context of random quantum circuits. But it is sometimes useful at some places which allow it, for eg, \cite{dawson} uses it to prove $\textbf{BQP} \subseteq \textbf{PP}$ as it simplifies by providing a straightforward view of this result.\\
In our setting of two-qubit Haar random circuits, as of yet it's really difficult to make much use of sum-over-paths technique, however, as we will see, we can certainly get motivated by certain aspects of it. By drawing a path diagram \ref{fig3_feynmanrandomckts}, it is straightforward to agree on two things - First, in the initial stages, as the two-qubit haar random gates are applied at each time step the overall state starts to become more entangled than the previous one (with high probability, since the set of random unitary gates that can effect the entanglement of input state is quite large compared to those who can't). Second, there is an upper bound on how much entanglement can a unitary \textit{induce} (or \textit{remove}) in a state. For instance, given an unentangled state, say, $\ket{0}^{00}$ some of the most entangled states achievable by a two-qubit random gate are Bell states and no other two-qubit random gate can produce more entanglement than what is achievable by bell states. Hence, all possible entanglement changes by a unitary must be upper bounded.
\\
Having deduced what we can from the \textit{path-diagrams} we try to search for a rule that an \textit{optimal circuit} for target state $\ket{\psi}$ must follow. By \textit{optimal circuit} we imply a circuit generating the desired target state $\ket{\psi}$ with the least number of gates $r$, arranged in an architecture $A$. The complexity of $\ket{\psi}$ is hence given by $\mathscr{C}_{state}(\ket{\psi}) = r$. Notice, we talk of \textbf{an} \textit{optimal circuit} - as you can generate the target state $\ket{\psi}$ with a circuit by tweaking $A$ and selecting different two qubit Haar-random gates while keeping number of gates $r$ constant. \\
We are hence motivated to conjecture that the sum of state-entanglement change over the discrete unitary time-steps should be the minimum along \textit{an} \textit{optimal circuit}, with high probability.

\section{Background}

\subsection{Feynman Path-Integral Formalism in Circuits}
\begin{figure}
    \centering
    \includegraphics[width=0.5\linewidth]{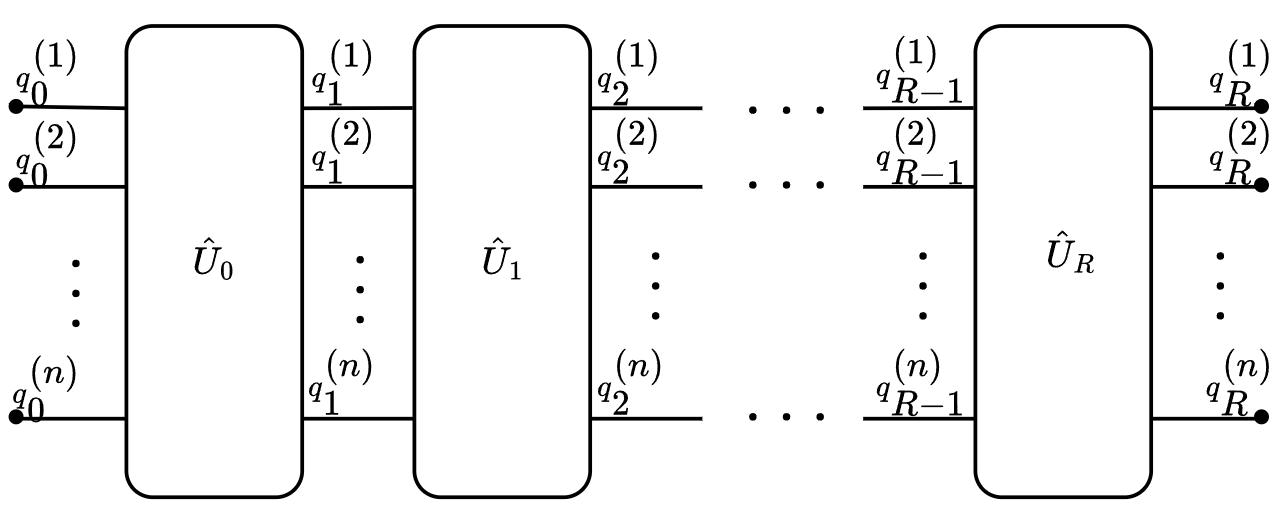}
    \caption{\small A circuit diagram with $R$ unitary gates illustrating the labeling convention.}
    \label{fig1_cktlabel}
\end{figure}

We provide a brief overview of \textit{feynman formalism} in context of discrete time systems such as quantum circuits. Suppose the configuration of the circuit is labeled by $q$, where configuration refers to the basis in which the quantum states of the circuit are expressed. The configuration of $n$ systems circuit (or $n$-qubit circuit) is given by a vector $\vec{q} \equiv (q^{(1)},...,q^{(n)})$, where $q^{(l)}$ is the configuration of $l$th system. Further, suppose that the circuit is composed of $R$ unitaries,  $\{\hat{U}_{i}\}_{i=1}^{r}$, so the circuit unitary becomes $\hat{U}=\hat{U}_R\hat{U}_{(R-1)}...\hat{U}_2\hat{U}_1$. Given the discrete nature of circuit dynamics we discretize time into R steps and configuration at time step $i$ is thus given by $\vec{q} \equiv (q^{(1)}_k,...,q^{(n)}_k)$ and a discrete-time path is a sequence of $R+1$ configurations, $\gamma=(\vec{q}_0,\vec{q}_1,...,\vec{q}_R)$.\ref{fig1_cktlabel} clears up the labeling convention for the discrete-time paths. The transition amplitude for $\ket{\vec{q}_0}$ -> $\ket{\vec{q}_R}$:
\begin{equation}
    \bra{\vec{q}_R}\hat{U}\ket{\vec{q}_0}=\sum_{\vec{q}_{R-1}\in(\mathbb{Z}_d)^n} ... \sum_{\vec{q}_{1}\in(\mathbb{Z}_d)^n}\prod_{i=1}^{R} \bra{\vec{q}_i}\hat{U}_i\ket{\vec{q}_{i-1}}
\end{equation}

The transition amplitude is the summation of all the amplitudes contributed by every path at target configuration $\vec{q}_R$. Amplitude for discrete-time path $\gamma \in (\mathbb{Z}_d^{n(R+1)})$ can be defined by,
\begin{equation}
    \bra{\vec{q}_R}\hat{U}\ket{\vec{q}_0} = \sum_{\gamma\in\Gamma(\vec{q}_0,\vec{q}_R)}A(\gamma)
\end{equation}
where $\Gamma(\vec{q}_0,\vec{q}_R)$ denotes the set of discrete-time paths from $\vec{q}_0$ to $\vec{q}_R$.\\
This functional over paths inside $\sum$ has a well known form:
\begin{equation}
    A(\gamma) = \mathcal{A}(\gamma)e^{iS[\gamma]}
\end{equation}
where $\mathcal{A}(\gamma)$ is the amplitude magnitude and $S[\gamma]$ is \textit{action functional/classical action} associated with the path $\gamma$.
\\
When dealing with physical systems in Quantum Field Theory and circuits whose all gates are \textit{balanced} the magnitude of the amplitude, $\mathcal{A}(\gamma)$ is path-independent and the phase $S[\gamma]$ is path-dependent. However, this is generally not the case with quantum circuits, especially for two-qubit Haar random quantum circuits - the ones we consider in our paper - where it's highly unlikely we'll obtain a circuit with all of its gates balanced.
\\
Although \cite{Penney_2017} tries to understand the phase $S[\gamma]$ as the action functional for a discrete-time path $\gamma$ in a classical counterpart of the quantum circuit. They do this for only special kind of circuits - CV Clifford Circuits and Discrete Clifford Circuits with \textit{"quopit"} configurations. There hasn't been any work that understands the action functional for general two-qubit Haar Random circuits. In this work, we do not directly address this question. Instead, we draw intuition from the random paths a circuit takes to prepare a state at a given time step $i$ (here $\vec{q}_i$) suggesting that the degree of entanglement in the state is influenced by the discrete-time trajectories that lead to its preparation.

\subsubsection{Circuit Analysis using Feynman Paths}
\label{feynmancktanalysis}
We provide insights into using \textit{Feynman paths} view for analyzing the circuit for Deutsch's algorithm for determining whether a function is balanced - a classic textbook problem. As we will demonstrate, this greatly simplifies the analysis and encourages us to interpret quantum circuits through this perspective, providing an intuitive understanding of the system as a whole.

In the \textit{Feynman paths} view of the circut, we layout the \textit{basis} of the circut in the y-axis and manipulate the amplitudes at each time step according to the unitaries. Repeating this for every gate in the circuit we can observe the interference of all paths to extract out what configurations/basis "survive" and contribute how much to the final state. In this way, without explicitly calculating the "action" of each unitary, we would have analyzed the circuit at hand.
The process is more clearly depicted in fig \ref{fig1_cktlabel}.\\
We employ the similar process in the next section that motivates our conjecture.
\begin{figure}
    \centering
    \includegraphics[width=1\linewidth]{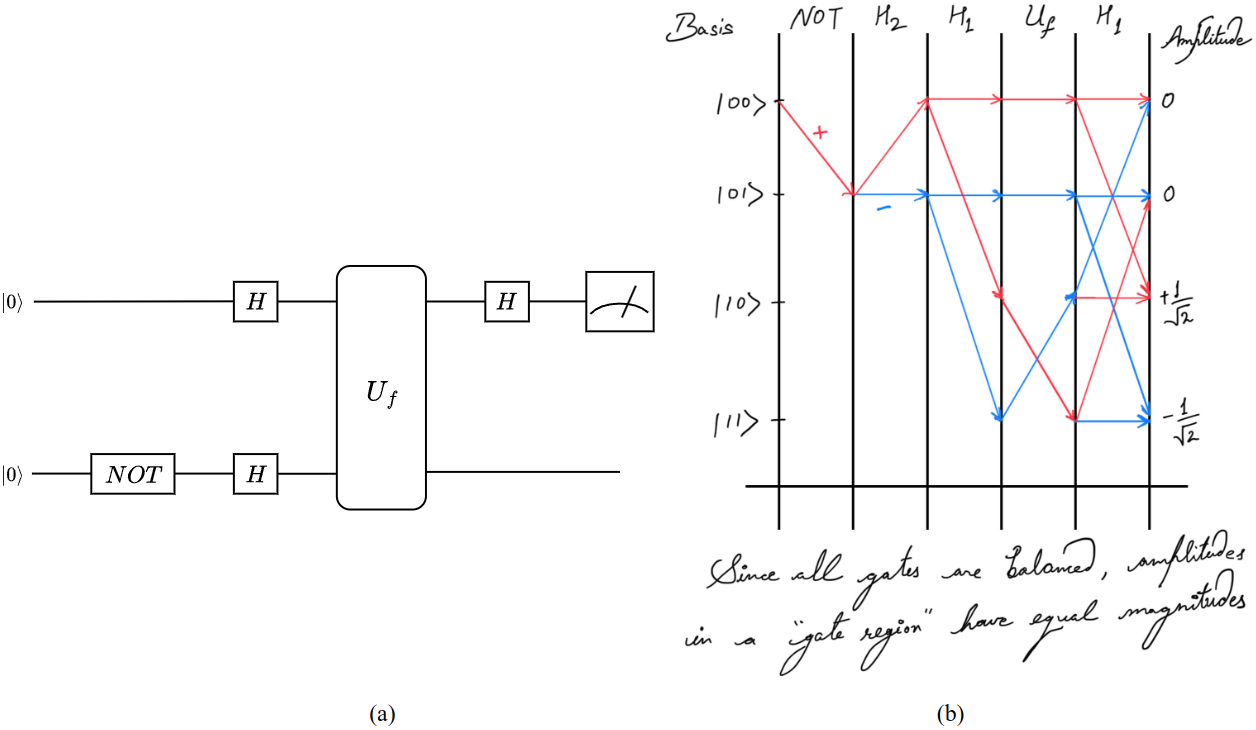}
    \caption{\small (a) Circuit diagram of Deutsch's algorithm, with the black-box function $f$ defined as $f(x) = \neg x$. (b) Visualization of the \textit{Feynman paths} perspective for Deutsch's algorithm, where positive amplitudes are shown in \textcolor{red}{red} and negative amplitudes in \textcolor{blue}{blue}.}
    \label{fig2_deutsch}
\end{figure}

\subsection{State Entanglement}
There are a lot of \textit{entanglement measures} \cite{duncan2024vonneumanns1927trilogy},\cite{entanglementmeasures2006survey},\cite{geometric_entanglement},\cite{negative_entanglement} in the literature of quantum mechanics and quantum cryptography for pure and mixed states. We list two such measures but we'll largely assume for the matter of this paper that \textit{pure} state entanglement has been quantified by geometric entanglement.
\subsubsection{Von Neumann Entropy:}
For a pure state $\ket{\psi}_{AB}$ of a bipartite system, the \textit{Von Neumann Entropy} of the reduced density matrix $\rho_A = Tr_B(\ket{\psi}\bra{\psi})$ of a subsystem $A$ is a direct measure of entanglement:
\begin{equation}
    S(\rho_A) = -Tr(\rho_A log\rho_A)
\end{equation}
This entropy quantifies the uncertainty or \textit{mixedness} of a subsystem $A$ with subsystem $B$. Similarly, the entanglement of the pure state $\ket{\psi}$ can be captured by the Von Neumann entropy of subsystem B.

\subsubsection{Relative Entropy of Entanglement:}
It is defined as the relative entropy between a given quantum state $\rho$ and the closest separable state $\sigma_{sep}$,
\begin{equation}
    E_R (\rho) = S(\rho||\sigma_{sep})
\end{equation}
where $S(\rho||\sigma_{sep}$ is the relative entropy given by $S(\rho||\sigma) = Tr(\rho log\rho - \rho log\sigma)$. This measure provides an operational definition of entanglement, quantifying the "distance" between the state and the set of separable states.

\subsubsection{Geometric Entanglement:}
\label{geoent}
\textit{Geometric entanglement} measures how far a quantum state is from the set of separable (unentangled) states. For a pure state $\ket{\psi}$,
\begin{equation}
    E_G(\ket{\psi}) = 1 - \max\limits_{\ket{\phi}\in\mathcal{S}}|\braket{\phi}{\psi}|^2,
\end{equation}
where $\mathcal{S}$ is the set of separable states, and the quantity $\max\limits_{\ket{\phi}\in\mathcal{S}}|\braket{\phi}{\psi}|$ represents the maximum overlap between the state and the closest separable state. This measure is particularly useful in multipartite quantum systems and thus, we suggest considering state entanglement through this measure for our further discussions on entanglement.

\section{Can Entanglement lead us to 'an' optimal circuit?}
This section is organized as follows. First, we introduce the setup of our random quantum circuit and definitions. Second, we argue how \textit{feynman paths} view motivated us to think of circuit optimization problem in terms of entanglement minimization of some sorts. Third, we introduce our measure of \textit{state-path-entanglement}, a sum of entanglement changes over the whole \textit{state-path}. The formal definition of a \textit{state-path} is provided later in this section, just before the conjecture is introduced. In the conclusion, we encourage considering quantum algorithms and problems in quantum complexity through the \textit{Feynman paths} viewpoint and suggest that \textit{Minimum entanglement-path conjecture} might have significant implications for circuit optimization and other related questions in quantum complexity theory.

\subsection{Preliminaries}
We formalize the notion of a \textit{random quantum circuit} and \textit{circuit/state complexity} for the sake of completion. This paper concerns a system of n qubits. Let $U_{j,k}$ denote a unitary gate that operates on qubits j and k. Such gates are general in the sense that they need not be geometrically local. An \textit{architecture} is an arrangement of a fixed number of gates, $R$.
\\
\small
\textbf{Definition 1} \textit{(Architecture)}: \textit{An architecture is a directed acyclic graph containing $R\in\mathbf{Z}_{>0}$ vertices (gates). Two edges (qubits) enter each vertex and two edges exit.}\\
\textbf{Definition 2} \textit{(Random quantum circuit)}: \textit{Let $\mathbb{A}$ denote an arbitrary architecture. A probability distribution can be induced over the architecture-$\mathbb{A}$ circuits as follows: For each vertex in $\mathbb{A}$, draw a gate Haar-randomly from SU(4). Then, contract the unitaries along the edges of $\mathbb{A}$. Each circuit so constructed is called a random quantum circuit.}\\
\textbf{Definition 3} \textit{(Exact Circuit/State Complexities)}: Let \( U \in SU(2^n) \) represent an \( n \)-qubit unitary. The \textit{(exact) circuit complexity} \( C_u(U) \) is defined as the minimum number of two-qubit gates required to implement \( U \) in a circuit. Similarly, for a pure quantum state \( |\psi \rangle \), the \textit{exact state complexity} \( C_{\text{state}}(|\psi \rangle) \) is the minimum number \( r \) of two-qubit gates \( U_1, U_2, \dots, U_r \), arranged in any architecture, such that
\[
U_1 U_2 \dots U_r |0^n \rangle = |\psi \rangle.
\]
\normalsize

\subsection{Feynman Paths for Random quantum circuits}

\begin{figure}
    \centering
    \includegraphics[width=0.5\linewidth]{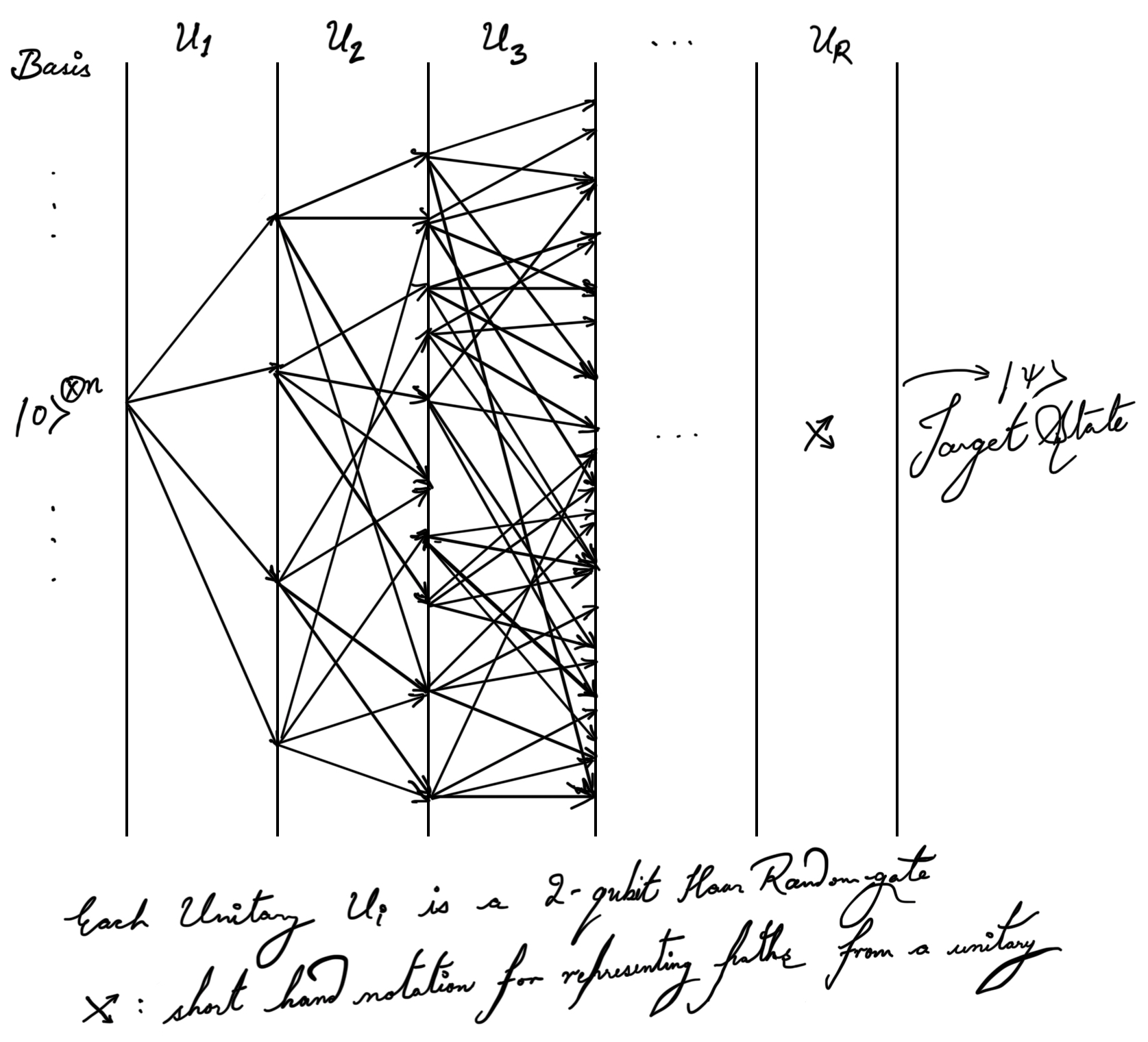}
    \caption{\small Feynman-path diagram for a Random quantum circuit.}
    \label{fig3_feynmanrandomckts}
\end{figure}

We repeat the same process as explained in Sec \ref{feynmancktanalysis} for Deutsch's algorithm. Since each time step $k$ consists of two-qubit Haar random gates, each \textit{"path-wave"} incident on a gate is split into four branches depending upon the inner working of the gate itself (which is Haar-random), which interfere to prepare the state $\ket{\vec{q}_k}$. We construct a corresponding path diagram for this random quantum circuit (Fig \ref{fig3_feynmanrandomckts}), providing an intuitive way to examine circuit optimization through the lens of state entanglement at each time step.

\subsection{Conjecture}
\begin{figure}
    \centering
    \includegraphics[width=0.6\linewidth]{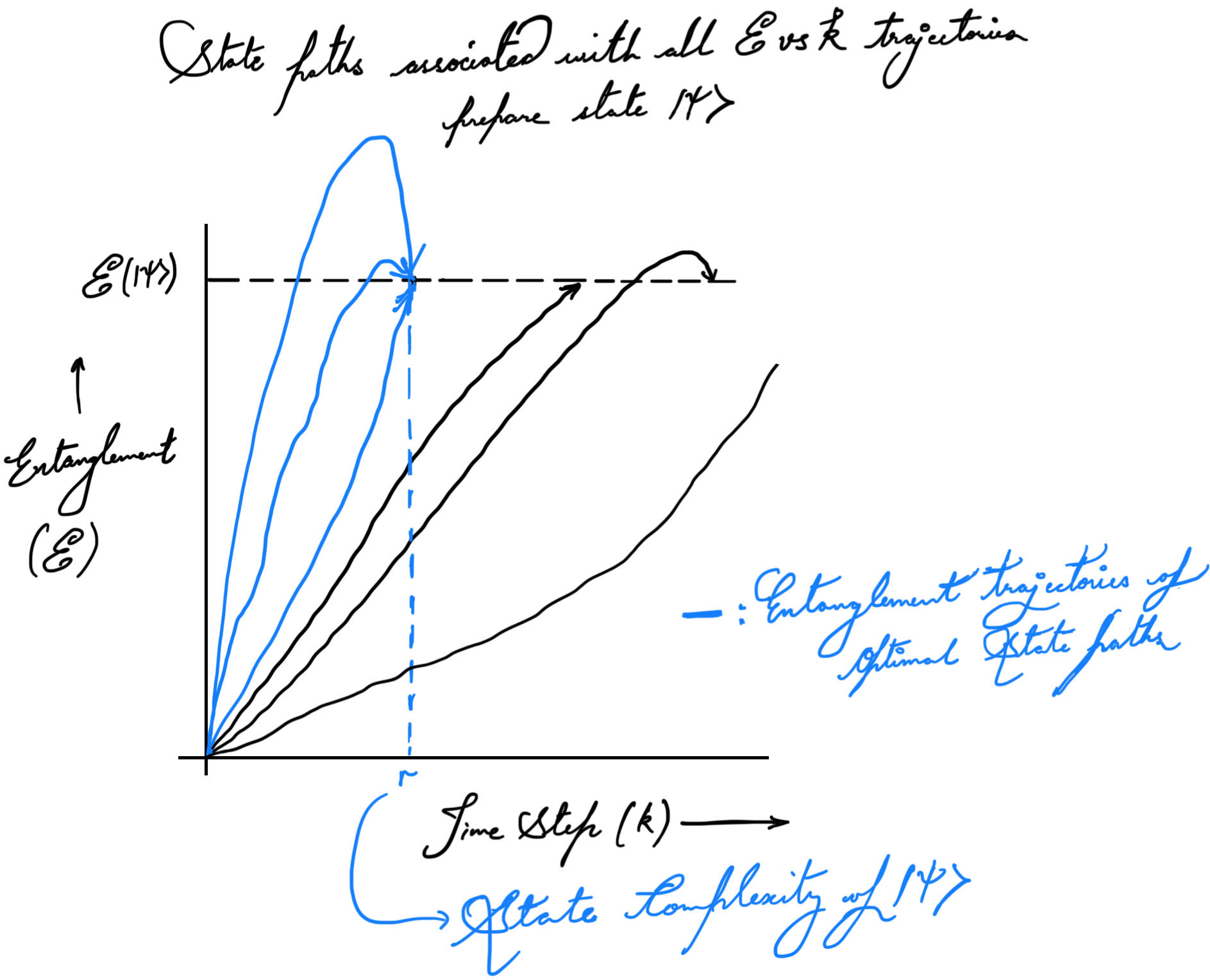}
    \caption{\small Entanglement ($\mathcal{E}$) vs. Time Step ($k$) plot showing the state entanglement evolution at each time step. Only the trajectories of \textit{state-paths} that prepare the state $\ket{\psi}$ are considered. Although the paths appear smooth, the discrete nature of circuits implies a staircase-like behavior. Each unit on the x-axis corresponds to one two-qubit Haar random unitary.}
    \label{fig4_ent_vs_T}
\end{figure}

We state our conjecture on the family of circuits with the spirit of finding at least one such circuit. Although the conjecture does not guarantee finding an optimal circuit design, it significantly narrows the search space for circuits by utilizing a circuit dependent entanglement measure.
Before that, we introduce a bit of terminology that will be used throughout this paper and found to be convenient. \\
\textbf{Definition 4} \textit{(State-path)}: \textit{A state-path ($\mathcal{P}$) is defined to be an ordered set of states $\{\ket{\psi_k}\}_{k=1}^{R}$ prepared by a circuit at each time step k. A state-path uniquely associated to a circuit $\mathrm{C}$ is denoted as $\mathcal{P}_{\mathrm{C}}$}.
\\
We introduce Entanglement($\mathcal{E}$) vs Time step($k$) plot \ref{fig4_ent_vs_T}. Entanglement($\mathcal{E}$) is plotted on the y-axis and Geometric entanglement measure \ref{geoent} is used to measure entanglement at each time step. The x-axis is evenly divided into intervals of \textit{"time"} corresponding to each two-qubit Haar random unitary. The greater the \textit{entanglement-trajectory} is spread out in x-component, the more is the complexity of the circuit (Note that this is different from the target state complexity as an optimal circuit might exist that prepares the target state).
\\\\\\
\textbf{Definition 5} \textit{(Path-entanglement sum)}: For a given state-path $\mathcal{P}_{\mathrm{C}}$ path-entanglement sum is defined as the summation of absolute changes in the entanglement $\mathcal{E}$ at every time step $k$.\\
\begin{equation}
    \textfrak{E} = \sum\limits_{k=1}^{R} |\Delta\mathcal{E}_k|,
\end{equation}

To achieve the entanglement embodied in a target state $\ket{\psi}$ there are a lot of other states which might share the same entanglement measure. Similarly, to prepare $\ket{\psi}$ there can a lot of different state-paths but only a certain set of these states would be made up of an optimal number of gates, $r$ (which would essentially be the \textit{state complexity} of $\ket{\psi}$). Hence it becomes essential to formalize the notation further to avoid any confusion.\\
Let $\mathscr{F}_{\ket{\psi}}$ denote the family of all \textit{state-paths} preparing state $\ket{\psi}$. Further, $\mathscr{F}_{\ket{\psi}}^{\textfrak{E}} \subseteq \mathscr{F}_{\ket{\psi}}$, where the former is the family of all \textit{state-paths} preparing $\ket{\psi}$ and have \textit{Path-entanglement sum} as $\textfrak{E}$.
\\
\textbf{Minimum entanglement-path conjecture:} An \textit{optimal state-path} is a member of the family $\mathscr{F}_{\ket{\psi}}^{\textfrak{E}_{min}}:=\min\limits_{\textfrak{E}}\mathscr{F}_{\ket{\psi}}^{\textfrak{E}}$ with high probability. That is, 
\begin{equation}
\mathbb{P} \left( \ket{\psi} \in \mathscr{F}_{\ket{\psi}}^{\textfrak{E}_{\text{min}}} \right) \geq 1 - \epsilon,
\end{equation}
where $\epsilon>0$ and $\mathscr{F}_{\ket{\psi}}^{\textfrak{E}_{min}}$ is the family of \textit{state-paths} with target state $\ket{\psi}$ with \textit{path-entanglement sum} being the minimum possible $\textfrak{E}_{min}$.
\\
A corollary directly follows from this conjecture: \\
\textbf{Corollary:} Let $\textfrak{P}\subseteq\mathcal{F}_{\ket{\psi}}$ be a set of \textit{state-paths} preparing $\ket{\psi}$. If $\exists\mathcal{P},\mathcal{P'}\in\textfrak{P}$ such that $\mathcal{P}\in\mathscr{F}_{\ket{\psi}}^{\textfrak{E}_{min}}$,
$\mathcal{P'}\notin\mathscr{F}_{\ket{\psi}}^{\textfrak{E}_{min}}$, then $\mathcal{P}$ is "closer" to an optimal path than $\mathcal{P}'$, \textit{with high probability}.

\section{Discussion}
This conjecture arose from the intuitive motivations from \textit{feynman's path view} which has long been established and has been really successful in solving problems and providing a more intuitive understanding of relativistic and non-relativistic quantum mechanics alike. Although much of the success of this perspective has been seen for the situations where the magnitude of the amplitude $\mathcal{A}$ is path independent -  this is generally not the case for quantum circuits. Furthermore, as of yet we still don't have a description of \textit{action functional} $S[\gamma]$, in the context of quantum random circuits. In spite of all that, this view of looking at quantum circuits still motivates us to think about state-entanglement evolution at each time step for an optimal circuit family and try to look for a general rule on how we can optimize a given circuit design with the target state $\ket{\psi}$.
\textit{Minimum entanglement-path conjecture} allows us to search for an optimal circuit design over a much smaller space. This space reduction could prove valuable for quantum circuit optimization algorithms in terms of faster convergence and other related works in quantum complexity theory.
However, in some "clever" situations that make up a small set of the vast space of random quantum circuit family, for example, when the entanglement entropy of target state $\ket{\psi}$ is $0$, this conjecture might not be of much use in reducing the search space for an optimal \textit{state-path}. But for a vast majority of target states, this conjecture should provide a significant reduction over the \textit{state-path} search space with high probability.



\section{Conclusion \& Future Work}
Inspired from \textit{Feynman's path formalism} in quantum theory, we try to provide a conjecture based on \textit{entanglement evolution} throughout the circuit for optimizing circuit design that prepare a given target state $\ket{\psi}$. 
While \textit{feynman's path view} for random quantum circuits has limited use in rigorous analyses — mostly due to the lack of a well-defined action functional in this a general circuit context, it still serves as a heuristic for attempting to almost solve the problem in terms of entanglement evolution. Some related works \cite{xu2021lagrangianformalismquantumcomputation}, \cite{Penney_2017} have suggested relations between the \textit{action functional} and circuit complexity. While similar relationship has been established in the context of adiabatic quantum computing, it is still not clear in discrete random quantum circuits. 
We strongly believe that developing a rigorous foundation for the \textit{feynman's path view} in random quantum circuits could yield more insights into this problem and other related areas in quantum complexity theory. \\
While the conjecture doesn't guarantee always to provide us with a circuit design that correctly approximates \textit{an} optimal circuit, a promising likelihood is suggested - though we do not quantify this likelihood rigorously. If the conjecture holds true with small probability of failing, it might be useful in circuit optimization and complexity theory research. Future work of the author will aim at establishing a probability bound for the conjecture's validity and to rigorously formalize the \textit{feynman's path view} for general quantum circuits.



\bibliographystyle{splncs04}
\bibliography{main}







\end{document}